# Strong Coulomb drag and broken symmetry in double-layer graphene

R. V. Gorbachev[1], A. K. Geim[1,2], M. I. Katsnelson[3], K. S. Novoselov[2], T. Tudorovskiy[3], I. V. Grigorieva[1,2], A. H. MacDonald[4], K. Watanabe[5], T. Taniguchi[5], L. A. Ponomarenko[1,2*]

[1]*Manchester Centre for Mesoscience & Nanotechnology, University of Manchester, Manchester M13 9PL, UK*

[2]*School of Physics & Astronomy, University of Manchester, Manchester M13 9PL, UK*

[3]*Institute for Molecules and Materials, Radboud University of Nijmegen, 6525 AJ Nijmegen, The Netherlands*

[4]*Department of Physics, University of Texas at Austin, Austin TX 78712, USA*

[5]*National Institute for Materials Science, 1-1 Namiki, Tsukuba, 305-0044 Japan*

**Spatially separated electron systems remain strongly coupled by electron-electron interactions even when they cannot exchange particles, provided that the layer separation $d$ is comparable to a characteristic distance $l$ between charge carriers within layers. One of the consequences of this remote coupling is a phenomenon called Coulomb drag, in which an electric current passed through one of the layers causes frictional charge flow in the other layer. Previously, only the regime of weak ($d \gg l$) to intermediate ($d \sim l$) coupling could be studied experimentally. Here we use graphene-BN heterostructures with $d$ down to 1 nm to probe interlayer interactions and Coulomb drag in the limit $d \ll l$ where the two Dirac liquids effectively nest within the same plane, but still can be tuned and measured independently. The strongly interacting regime reveals many unpredicted features that are qualitatively different from those observed previously. In particular, although drag vanishes because of electron-hole symmetry when either layer is neutral, we find that drag is at its strongest when *both* layers are neutral. Under this circumstance, drag is positive in zero magnetic field but changes its sign and rapidly grows in strength with field. The drag remains strong at room temperature. The broken electron-hole symmetry is attributed to mutual polarization of closely-spaced interacting layers. The two-fluid Dirac system offers a large parameter space for further investigation of strong interlayer interaction phenomena.**

Since the first observation of Coulomb drag in GaAlAs heterostructures [1-3], double-layer electronic systems have been attracting unwavering interest (for review, see [4,5]). A number of new interaction phenomena have been reported, which appear due to many-body ground states formed jointly by charge carriers from both layers. These phenomena include even-denominator fractional quantum Hall states [6-8] and interlayer excitonic superfluidity [4,9-13]. It is believed that stronger interactions could lead to excitonic superfluidity at higher temperatures ($T$) and open new avenues for fundamental research and applications [14,15].

Until recently, GaAlAs heterostructures with two quantum wells separated by a relatively thin tunnel barrier (effective $d$ down to 15 nm) were the only experimental systems allowing reliable measurements of Coulomb drag [1-13,16]. The advent of graphene boron-nitride heterostructures offers a double-layer system with stronger interlayer interactions [17,18]. First, charge carriers in graphene are confined within a single atomic plane and a few atomic layers of hexagonal boron-nitride (hBN) are sufficient to isolate graphene electrically [18-20]. This allows double-layer heterostructures with $d \sim 1$nm, an order of magnitude smaller than achieved previously. Second, the relatively small dielectric constant of hBN ($\varepsilon \approx 4$) also helps to increase the interaction strength. Third, charge carriers in graphene can be continuously tuned between electrons ($e$) and holes ($h$) from densities $n > 10^{12}$cm$^{-2}$ all the way through the neutral state where $l = 1/n^{1/2}$ nominally diverges. This makes it possible to access the limit $d/l \ll 1$ or, effectively, zero layer separation. Fourth, Coulomb drag in graphene heterostructures can be studied not only in the conventional Fermi liquid regime (where the Fermi energy $E_F$ is larger than $T$) but also in the little explored Boltzmann regime ($E_F < T$). These features suggest a unique venue for searching for new interaction phenomena. Accordingly, double-layer graphene has attracted significant theoretical interest and already become a subject of intense debates about predictions for both Coulomb drag

[21-27] and excitonic superfluidity [15,28]. Recent experiments [29] reported the observation of Coulomb drag in double-layer graphene in zero magnetic field $B$ in devices in which the layers were separated by a several nm thick alumina barrier and had charge carrier mobilities $\mu \sim 10,000$ cm$^2$/Vs.

In this paper, we describe strong Coulomb drag and its anomalous behavior in high-quality graphene-hBN heterostructures with mobilities $\mu \sim 100,000$ cm$^2$/Vs and $d$ down to 1 nm (trilayer hBN). In zero $B$ and away from the neutrality point (NP), we find the standard quadratic $T$ dependence, a hallmark of Coulomb drag between degenerate Fermi liquids. By varying $n$ and $T$, we observe a crossover between the Fermi liquid ($E_F >> T$) and interacting Boltzmann gas ($E_F < T$) regimes, which shows up as a clear maximum in drag at $E_F \approx 2T$. In the former regime, the dependences on $n$ and $d$ are much weaker than those found in GaAlAs heterostructures, and drag changes little for $d <$4nm indicating that the limit of zero $d$ is reached. These results generally agree with the theory developed recently for $d < l$ [25,26]. However, we find significant differences too. Most of them appear in the regime where both graphene layers are neutral. The neutral state unexpectedly exhibits the strongest drag. It is positive in zero $B$ but changes the sign and rapidly grows in relatively weak $B$ =0.1-1T. We explain the positive drag in terms of interaction-induced correlations between $e$-$h$ puddles in the two layers. The strong negative magneto-drag is attributed to interlayer exciton-like correlations between half-filled zero Landau levels (LLs).

**Devices and measurements**
The schematics of our experiments are shown in Fig. 1a. The devices consist of two graphene monolayers separated by a thin hBN spacer. This double-layer structure is encapsulated between relatively thick (>20nm) hBN crystals. The encapsulation is important to achieve high $\mu$ and little charge inhomogeneity $\delta n$ [17,18,30]. The entire heterostructure is assembled on top of an oxidized Si wafer that serves as a bottom electrode. Carrier densities $n_T$ and $n_B$ in the top and bottom layers, respectively, are controlled by voltages $V_T$ and $V_B$ applied to top and bottom electrodes. The two graphene layers are shaped into multiterminal Hall bars (width of 1-2 $\mu$m), which are aligned on top of each other. By measuring longitudinal and Hall resistivities ($\rho$ and $\rho_{xy}$, respectively) we can fully characterize each layer [17,18]. In particular, the positions of the NP as a function of $V_T$ and $V_B$ are identified by zeros in $\rho_{xy}$ (peaks in $\rho_{xx}$). Away from the NP, $\rho_{xy} = B/ne$ and the Hall measurements yield $n$ ($e$ is the electron charge). This allows us to find $n_{T,B}$ as a function of $V_T$ and $V_B$. Details about our devices' fabrication and characterization can be found in Supplementary Information and in refs. [17,18,30].

Drag measurements were carried out by applying current $I_{drive}$ through one graphene layer, and measuring the induced voltage $V_{drag}$ in the second layer. The observed linear response $V_{drag} \propto I_{drive}$ allows us to present experimental data in terms of the drag resistivity $\rho_{drag}$ that is found to scale in the conventional manner with width and length of the Hall bars. Typically, the top layer has lower quality ($\mu_T \approx 30,000$ to 90,000 cm$^2$/Vs) and we employ it as drive. Higher-quality bottom graphene ($\mu_B \approx 50,000$ to 120,000 cm$^2$/Vs) serves as the drag layer. If the drive and drag layers are interchanged, $\rho_{drag}$ does not change, that is, the Onsager relation holds.

For hBN spacers down to 3 atomic layers in thickness, the tunneling resistance is sufficiently high (>1MOhm at low bias) and independent of $T$ [18-20] to ensure a negligible contribution of interlayer tunneling to the measured voltage signal (Supplementary Information). The interlayer resistance becomes unacceptably low for bilayer hBN (~10 kOhm). This implies that $d$ =1nm is the minimum separation that can realistically be achieved in drag experiments, even using quality insulators like hBN [20]. In some measurements, we have applied interlayer voltage $V_{int}$ which conveniently results in equal doping $n_T = -n_B$ (positive and negative $n$ are assigned to $e$ and $h$, respectively). However, the latter approach is impractical for $d <$4nm because the tunnel current increases with bias, effectively limiting the electrical doping achievable by varying $V_{int}$ to <10$^{11}$ cm$^{-2}$ for small $d$.

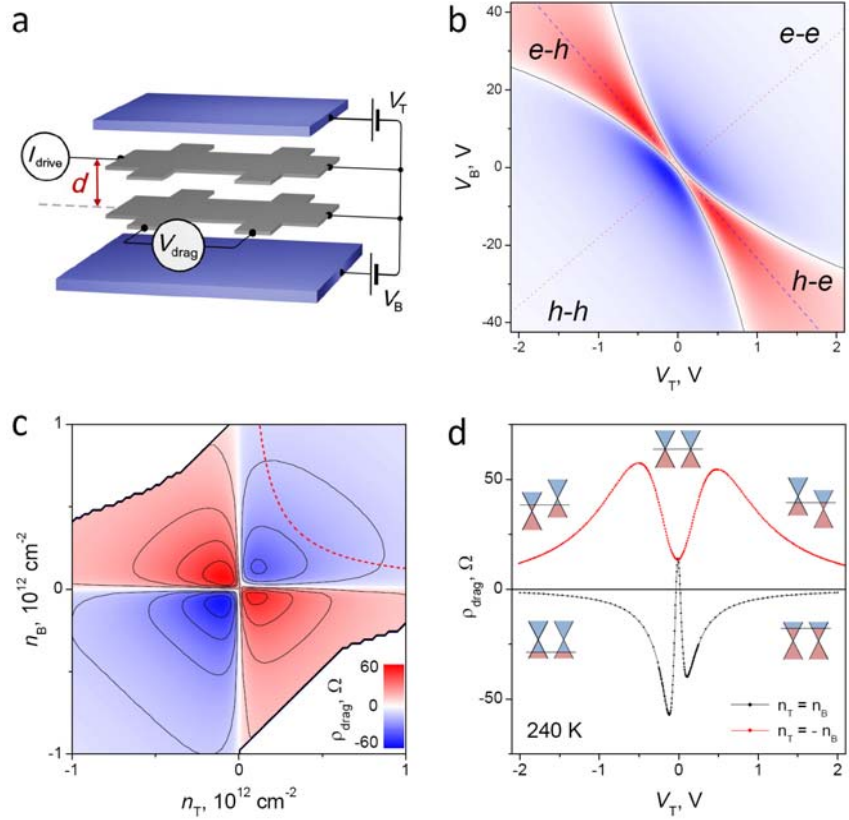

Figure 1. **Coulomb drag in double-layer graphene heterostructures in zero magnetic field**. **a** – Schematics of our devices and measurements. **b** – $\rho_{drag}$ as a function of $V_T$ and $V_B$ for a device with a trilayer hBN spacer ($d \approx 1$nm). The black curves indicate zero-lines of $\rho_{drag}$; the color scale is $\pm 60\Omega$ [see the inset in (c)]. $T$ =240K: in most cases, this was our highest $T$, chosen to be sufficiently close to room $T$ and, at the same time, to prevent device damage that often occurred above 240K. The device exhibits little chemical doping (<<$10^{11}$ cm$^{-2}$) and, for clarity, $V_T$ and $V_B$ are offset to zero at the dual NP. **c** – $\rho_{drag}$ from (b) replotted as a function of $n_T$ and $n_B$. The solid curves are iso-levels at every 12$\Omega$. The blank regions separated by thick lines contain no data. The dashed curve illustrates the functional dependence expected in the weakly interacting regime, $\rho_{drag} = f(n_T \times n_B)$. **d** – Behavior of $\rho_{drag}$ under the equal density conditions, $n_T = -n_B$ and $n_T = n_B$, which are indicated in (b) by the dashed and dotted lines, respectively. $\rho_{drag}$ in (d) is plotted as a function of $V_T$; $V_B$ is varied so that the densities are kept equal, allowing a common $x$-axis for (b) and (d). The Dirac-cone diagrams illustrate various doping regimes.

## Coulomb drag in zero magnetic field

Figure 1b shows $\rho_{drag}$ as a function of $V_T$ and $V_B$ for one of our thinnest devices. There are 4 distinct segments, which correspond to different combinations of electrons and holes in the two layers. If both layers contain carriers of the same (opposite) sign, $\rho_{drag}$ is negative (positive) [1-13,21-27]. The absolute value of $\rho_{drag}$ exhibits a maximum in each of the 4 segments at low carrier densities. The condition of zero drag is shown by the solid curves, which (closely but not exactly) follow the NPs. The pronounced asymmetry between the size of the red and blue segments is due to a large quantum capacitance contribution for small $d$, which results in strongly nonlinear functions $n_{T,B}(V_{T,B})$ [17,31,32]. These were determined experimentally and modeled theoretically, which allows us to replot the data from Fig. 1b as a function of $n_T$ and $n_B$. In the latter representation (Fig. 1c), we recover nearly perfect 4 quadrant symmetry.

To get a closer look at the behavior of Coulomb drag, we reduce the available parameter space to the symmetric situation in which $e$ and $h$ densities in both layers are equal. This matching density regime has been attracting particular attention [3-15]. The conditions $n_T = \pm n_B$ are marked in Fig. 1b by the dashed and dotted lines. Figure 1d shows changes in $\rho_{drag}$ along these lines. One can see that, for $e$-$h$ drag, $\rho_{drag}$ exhibits a relatively simple, double-humped behavior, which is easy to understand: Drag has to decrease with increasing $n$ as well as to disappear in neutral graphene. This necessitates maxima in $\rho_{drag}$ at a finite $|n|$. In the case of $e$-$e$ and $h$-$h$ drag, $\rho_{drag}$ has the opposite sign but exhibits roughly the same double-humped shape, except for quicker response to $V_T$ (cf. Fig. 1b-c). The major difference between the two curves in Fig. 1d appears near the dual NP where $\rho_{drag}(n_T = n_B)$ changes its sign to reach the same positive value as $\rho_{drag}(n_T = -n_B)$. The region of positive drag at the dual NP is also seen as a narrow gap between the curves marking zero $\rho_{drag}$ in Fig. 1b. Positive drag at the dual NP has been observed at $T > 50$K and zero $B$ for all our devices. This observation implies broken $e$-$h$ symmetry.

**Strongly interacting Dirac fermion liquids: comparison of experiment and theory**
Let us first examine drag's behavior away from the NP, in the Fermi liquid regime studied extensively by theory. We focus below on the case of equal $e$-$h$ densities ($n_T = -n_B \equiv n$), chosen for both simplicity and the theoretical interest it attracts. By measuring curves such as shown in Fig. 1d at different $T$, we have found that $\rho_{drag} \propto T^2$ for sufficiently high $n$ ($E_F > 3T$). Examples of this behavior are shown in Fig. 2a and in Supplementary Information. The $T^2$ dependence is expected at low temperatures for any interaction strength $d/l$ between graphene layers [21-27], provided that the electrons form a Fermi liquid state.

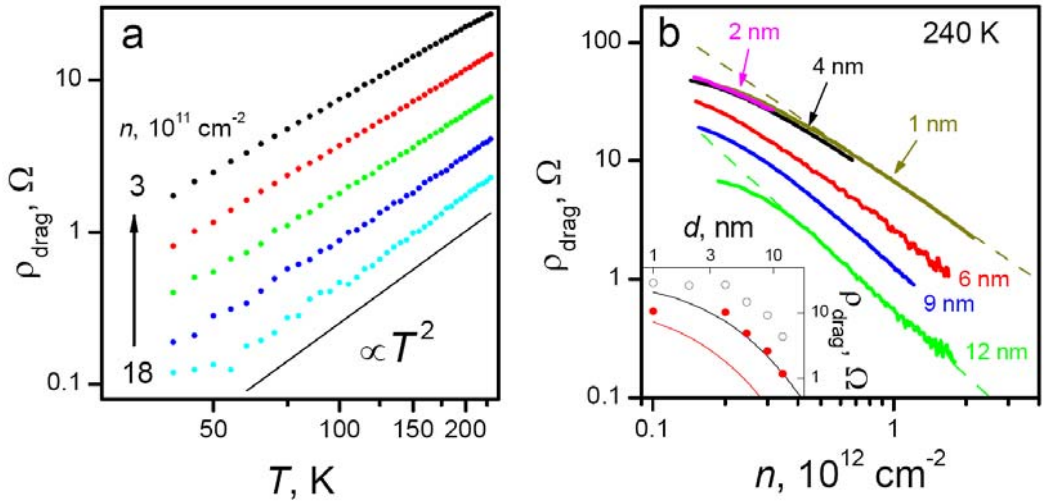

Figure 2. **Functional dependences of Coulomb drag away from the NP.** a – $\rho_{drag}$ as a function of $T$ for various $n$ between 3 and 18 $\times 10^{11}$ cm$^{-2}$; $d \approx 6$nm; $B = 0$. The solid line shows the expected functional dependence. Our devices are several microns in size and, at low $T$, exhibit strong mesoscopic (interference) fluctuations, which are observed in both drag and resistivity [32,33]. This limits the $T$ range in which quantitative comparison with the theory is possible to $T > 40$ K. To investigate Coulomb drag at lower $T$ would require much larger devices. b – $\rho_{drag}$ as a function of $n$ and $d$. The dashed lines are fits to $1/n^\alpha$ for $d = 1$ and 12 nm. The inset shows $\rho_{drag}$ as a function of $d$ for $n = 3$ and 7 $\times 10^{11}$ cm$^{-2}$ (open and closed symbols, respectively). The solid curves are theory plots for these two $n$ in the corresponding colors. $T = 240$K, but the functional forms hold over the entire $T$ range as witnessed by the parallel shift of the curves in (a).

As for the $d$ dependence, we find that for large $d > 4$nm $\rho_{drag}$ varies approximately as $1/d^2$ (Fig. 2b), that is, much slower than $1/d^4$ expected in the weakly interacting regime. Moreover, our devices show practically no increase in drag below 4 nm, in qualitative agreement with theory [25,26] that predicts $\rho_{drag}(d)$ to saturate at small $d$ (Fig. 2b). Another significant difference with respect to weakly interacting systems is the observed functional dependence. In the $d >l$ case, theory expects $\rho_{drag}$ to be a function $f$ of the product $n_B \times n_T$ [21-27] which should have resulted in strongly curved iso-levels such as shown by the dashed curve in Fig. 1c. Instead, we observe flat iso-levels which imply that $\rho_{drag}$ is closely described by the functional dependence $f(n_B+n_T)$. This behavior is robust and exemplified further in Supplementary Information. The theory [25,26] predicts a change in the functional dependence for $d <<l$ but cannot explain the observed $f$ (Supplementary Information). At $T$ >100K we find little variation in the measured $\rho_{drag}$ for different contacts and devices with the same $d$. Drag also changes little after thermal cycling or annealing, although these procedures often change $\mu$ and occasionally shift the NP. The high reproducibility and the well-defined $d$ (cf. GaAlAs heterostructures) provide us with confidence in comparing the absolute value of $\rho_{drag}$. The inset in Fig. 2b and Supplementary Information shows that the measured drag exceeds the theoretical values (calculated according to [26]) by a factor of $\approx 3$.

**Anomalous drag in neutral graphene in zero and finite $B$**

Now we focus on drag close to the dual NP. In zero $B$ and at high $T$, all our devices have exhibited positive drag at the dual NP. Moreover, this positive drag often develops into a sharp peak with decreasing $T$. As an example, Fig. 3a shows $T$ dependence of $\rho_{drag}(n)$ for a device in which the peak was particularly pronounced. At 240 K, we observe the same double-humped behavior as in Fig. 1d. At lower $T$, drag at the dual NP first grows, which is opposite to the $T$ dependence at finite $n$ (Fig. 2). As a result, the curves become triple-humped below 150 K and the central peak dominates the low $T$ curves, until drag eventually disappears under random-sign mesoscopic fluctuations [32,33].

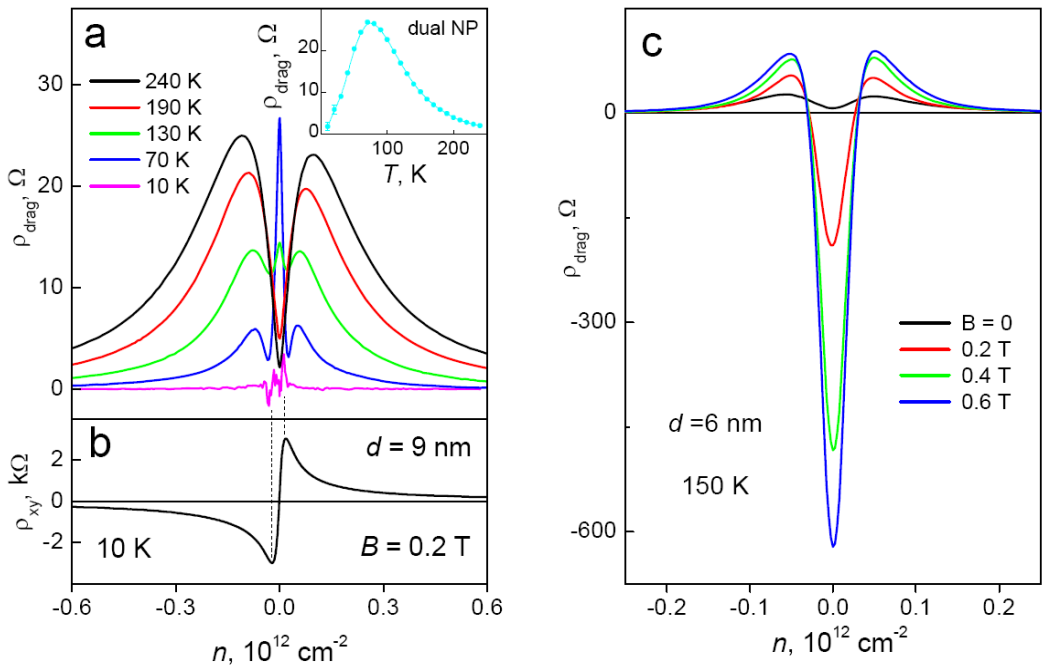

Figure 3. **Broken symmetry in neutral double-layer graphene. a** – $T$ dependence of $\rho_{drag}(n_T = -n_B \equiv n)$ in one of our devices in zero $B$; $d$ =9nm. A sharp peak emerges at the dual NP with decreasing $T$. Its $T$ dependence is shown in the inset. The outside maxima in $\rho_{drag}(n)$ shift approximately linearly with $T$ above 100 K. Their positions correspond to $E_F \approx 2T$, in agreement with ref. [25] that extends the drag theory into the Boltzmann regime. **b** – The regime of $e$-$h$ puddles can be seen on $\rho_{xy}(n)$ curves as a transitional region indicated by the dashed lines. **c** – Typical changes in $\rho_{drag}$ upon applying $B$. Magneto-drag at zero $n$ exhibits a nonmonotonic $T$ dependence such that it decreases with both increasing and decreasing $T$. For $B$ <1T, strongest drag occurs between 120 and 200 K, depending on $B$ (Supplementary Information).

The peak's strength varied from sample to sample so that, in some devices, it required $T$ <70K to detect the central peak (Supplementary Information). The sample dependent onset occurs because the $T$ dependence at the dual NP is nonmonotonic (inset in Fig. 3a) whereas the general background varies as $\propto T^2$. Accordingly, central peaks with smaller strength require lower $T$ to emerge. It is important to note that the central peak always appeared in the regime of *e-h* puddles. Their presence in each layer can be monitored by Hall measurements (Fig. 3b). At low $T$, the two extrema in $\rho_{xy}(n)$ indicate the crossover between a uniform Fermi liquid ($\rho_{xy} \propto 1/n$) and electron transport with a varying amount of *e-h* puddles ($\rho_{xy} \propto n$). We emphasize that the largest peaks at the dual NP were observed in devices with the widest *e-h* puddle regions, that is, with the largest $\delta n$. Thermal annealing could sometimes reduce $\delta n$, which resulted in weaker central peaks. No correlations with $\mu$ have been noticed. These observations suggest that the peak at the dual NP in zero $B$ is due to *e-h* puddles.

In GaAlAs heterostructures, Coulomb drag was found to exhibit its most interesting behavior in the magnetic quantum limit where only the lowest LL remains occupied [4-11]. Accordingly, we have probed magneto-drag in our devices, focusing on this quantum limit that in graphene occurs at zero $n$. Away from the dual NP, $\rho_{drag}$ increases with increasing $B$ and oscillates but these changes are relatively modest (Fig. 3c; Supplementary Information). The major qualitative change is found in the neutral state, in which case even relatively small $B$ causes Coulomb drag to alter its sign. Figure 3c shows that this is an overwhelmingly strong effect such that $\rho_{drag}$ increases by a factor of >100 in $B$ <1T. In larger $B$ >10T, the magneto-drag can reach >10 kOhm at ~50K. However, drag's behavior in high $B$ is complex, mostly because the field opens energy gaps near the NP [34], and individual graphene layers become increasingly insulating with increasing $B$ and lowering $T$. Therefore, we have limited our present report to moderate $B$ <5T, in which case drag exhibits the behavior shown in Fig. 3b. It is highly reproducible from sample to sample and remains qualitatively the same for all our devices and only its magnitude changes with $d$ (see Supplementary Information).

**Possible origins of broken symmetry in neutral graphene**
In the case of zero $B$, the peak at the dual NP has already attracted two possible explanations [35,36]. One relies on third-order interaction contributions that can be dominant at the Dirac point [35]. This model does not capture the experimental fact that the central peak is related to *e-h* puddles. The second explanation attributes the peak to the interlayer energy transfer mediated by electron-electron interactions and explains many features in the observed behavior [36]. We offer another explanation. As discussed in Supplementary Information, if *e-h* puddles are due to an external electrostatic potential, their distribution in the two layers is correlated and, therefore, average drag at the dual NP should be negative. We suggest that *e-h* puddles in our high-$\mu$ graphene are predominantly due to strain that results in a pseudo-electrostatic potential [37]. Under this assumption, *e-h* puddles are formed independently, if the two layers are far apart. However, for small $d$, Coulomb interaction leads to mutual polarization, and *e* puddles in one layer lie predominantly on top of *h* puddles in the other layer (Supplementary Information). This model explains the positive peak and accounts qualitatively for its $T$ and $n$ dependences. It is also consistent with the observation that the peak is large for large $\delta n$.

As for strong negative magneto-drag at the NP, no theory has been developed so far. Nonetheless, the problem can be mapped onto the case of two half-filled LLs in GaAlAs heterostructures, which result in giant negative drag [4,6,9-11]. The associated interlayer correlated many-body state is usually described in terms of condensation of excitons that consist of an electron in one layer (filling factor $\nu$ =1/2) bound to a vacancy-like state in the other $\nu$ =1/2 layer [4]. In high $B$, the parameter that controls the interaction strength is $d/l_B$ where $l_B$ is the magnetic length. The observation of negative drag in GaAlAs heterostructures at the total filling factor $\nu_T$ =1/2+1/2 requires $d/l_B$ <2 [38] and values below 1 could not be achieved [4,6,9-11,38]. Due to the Dirac-like spectrum, graphene's lowest LL ($N$ =0) is at zero energy so that, at the dual NP, there are two half-filled LLs [39].

As in the GaAlAs case, electrons at $N=0$ LL in one graphene layer can then pair with vacancy-like states at the $N=0$ LL in the other layer. In GaAlAs devices, the observation of the excitonic drag requires modest $B \approx 1$-2T but $T <1$K. It is surprising that strong interlayer correlations can apparently survive in graphene up to room $T$. To align this observation with intuition, we note first that the typical $d/l_B$ in our experiment is ~0.1 (Fig. 3c), an order of magnitude smaller than in GaAlAs heterostructures. Secondly, our devices exhibit mobilities ~100,000 cm$^2$/Vs, practically independent of $T$. Therefore, the high field regime ($\mu B >1$), in which cyclotron orbits are formed and $d/l_B$ (rather than $d/l$) defines the interaction strength, is reached at $B >0.1$T. Thirdly, although the room-$T$ quantum Hall effect in graphene is observable for $B >20$T [40], the build-up of the density of states at zero LL occurs in much lower $B$, as observed in capacitance measurements at room $T$ [31]. We believe that, under our experimental conditions, nascent $N=0$ LLs assist the strong negative drag at the dual NP.

**Conclusions**

Double-layer graphene heterostructures with ultra-thin hBN spacers allow us to achieve unprecedentedly strong interactions between electrically isolated two-dimensional systems, especially in a magnetic field. In the Fermi liquid regime, the observed Coulomb drag is in agreement with theory, except for a factor of about 3 in absolute value and an unexpected functional dependence $\rho_{drag} = f(n_B + n_T)$ on layer densities. In our opinion, the most interesting and previously unexplored regime is when both layers are neutral. In this case, strong sign-changing drag suggests broken $e$-$h$ symmetries and new physics. In zero $B$, this can be explained by interaction-induced spatial correlations between $e$-$h$ puddles or interaction-induced energy transfer. In magnetic field, interlayer exciton-like correlations offer a convenient explanation for large drag and invite one to search for high-$T$ excitonic condensates. To investigate and understand double-layer graphene over a wide range of $T$ and especially in quantizing $B$, much further work is needed, perhaps along the lines of enquiry being pursued for GaAlAs heterostructures during the last two decades.

*Acknowledgements* – We thank L. Levitov, M. Titov and A. Castro Neto for helpful discussions. This work was supported by the Royal Society, the Körber Foundation, Engineering and Physical Sciences Research Council (UK), the Office of Naval Research and the Air Force Office of Scientific Research.

**Supplementary Information**

## 1. Fabrication and measurement procedures

Fabrication of our graphene-hBN multilayer heterostructures involved multiple steps, in which graphene and thin crystals of hBN were isolated, deposited on top of each other, shaped and electrically contacted. Each layer was prepared separately on a polymer coated wafer and transferred onto a target crystal with micron positioning accuracy. Then, the layer was annealed in order to remove chemical residues, before being covered with the next crystal. Both graphene layers were shaped by oxygen plasma etching into multiterminal Hall bars, with typically 6 to 10 contacts (Fig. S1). The above fabrication procedure was previously described in ref. [S1] and has now been extended to include an encapsulating hBN layer (~20 nm thick) to cover the entire stack. The encapsulating layer also served as a gate dielectric for the top gate as shown in Figure S1. The entire heterostructure was assembled on top of an oxidized Si wafer (usually 300 nm of $SiO_2$).

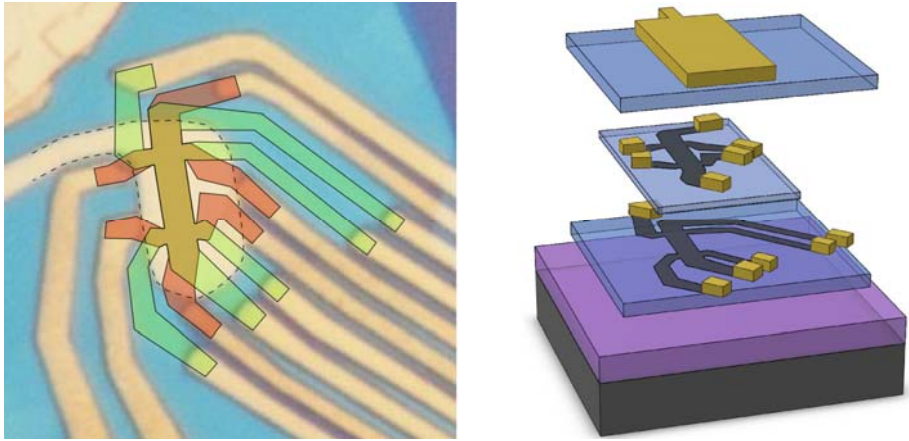

Figure S1. Optical micrograph of a double-layer graphene device (left) and its schematics (right). The layer sequence is as follows: oxidized Si wafer as a platform and back gate, hBN crystal as a substrate, bottom graphene Hall bar, ultra-thin hBN spacer, top graphene Hall bar, encapsulating hBN crystal and top gate. In the photo, the graphene Hall bars are shown in false colors (green and orange); the Au/Ti gate on top is outlined with the dashed curve. The spatial scale is given by the width of the Hall bars which is 1.5 µm.

Drag measurements were carried out using both DC and low-frequency lockin techniques, which yielded the same $\rho_{drag}$. Our preferred method was to use a DC current, then reverse its direction, take the average of two voltage signals and then repeat the procedure several times. This computer-automated procedure is equivalent to near-zero-frequency AC measurements.

Special care was taken to ensure that the leak current $I_L$ between top and bottom layers did not contribute into the measured voltage $V_{drag}$. Leakage was never a problem for $d \geq 2$ nm because of the high quality barrier provided by hBN [S2]. If $V_{int}$ was used to induce charge carriers ($n_T = -n_B$) we monitored $I_L$ as a function of $V_{int}$ and avoided $I_L$ above a few nA at typical $V_{int}$ ~1V. This translates into an interlayer resistance $R_{int}$ >100 MOhm, several orders of magnitude larger than in-plane graphene resistivity <10 kOhm. For $d$ =1 nm (3 hBN layers), $R_{int}$ was only ~1 MOhm in the limit of zero bias and decreased at higher $V_{int}$ [S2]. Therefore, we changed carrier densities in these devices only by gate voltages and did not use any bias $V_{int}$. If we applied a finite $V_{int}$ in devices with $d$ =1 nm, $I_L$ rapidly increased and we could observe changes in measured $V_{drag}$ due to $I_L$. This leakage contribution disappeared at small biases. In addition, we could monitor how much drag voltage was influenced by interlayer leakage by using different configurations of current contacts. This resulted in different distributions of leakage currents but at small $V_{int}$ the measured $\rho_{drag}$ showed little response. However, for a spacer made from bilayer hBN, $R_{int}$ was typically about 10 kOhm, and a significant portion of $I_{drive}$ flowed through the second layer. This makes the layers electrically coupled and disallows drag measurements for spacers less than 3 hBN layers. Note that if a gold film is used instead of graphene as one of the tunnel electrodes, $I_L$ increases by 2-3 orders of magnitude because of higher tunneling density of states [S2].

## 2. Additional examples of Coulomb drag's behavior in double-layer graphene

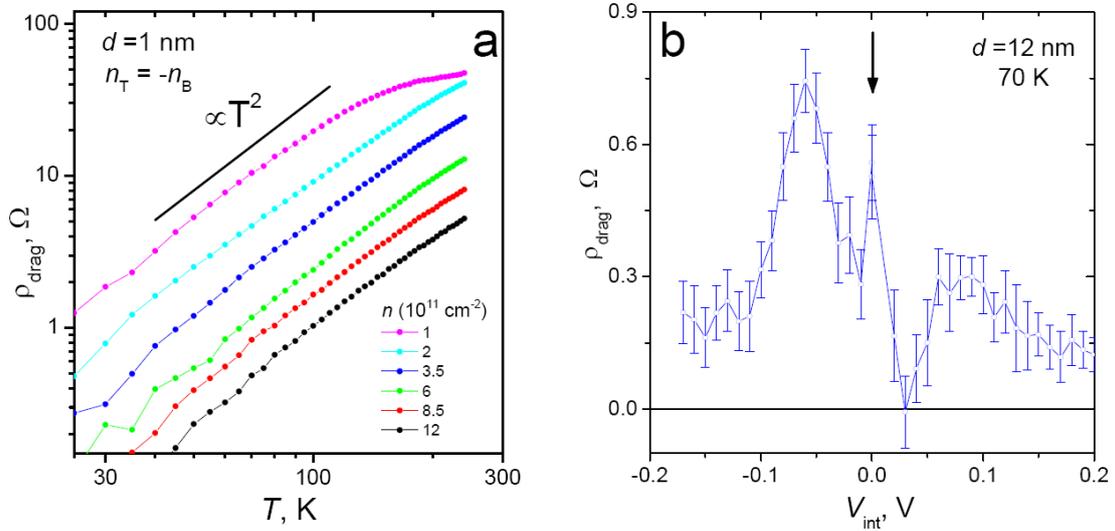

Figure S2. **a** – Universal $T$ dependence in the Fermi liquid regime and zero $B$. The same device as shown in Fig. 1 at various $n \equiv n_T = -n_B$ (cf. Fig. 2a). **b** – The positive peak at the dual NP is a universal feature, although in some devices it may require rather low $T$ for the peak to emerge. In (b), we change $n$ by interlayer bias $V_{int}$ that creates equal concentrations of $e$ and $h$ in the two layers ($n_T = -n_B$). The curve in (b) is strongly distorted because of mesoscopic fluctuations and the small magnitude of $\rho_{drag}$ for this relatively large $d$ =12nm. We are confident that the peak at zero $n$ is not mesoscopic because it is reproducible for various contacts and after thermal cycling, unlike fluctuations. The fluctuating background disappears at $T$ >120K, and the device shows positive drag at the dual NP, similar to all other devices.

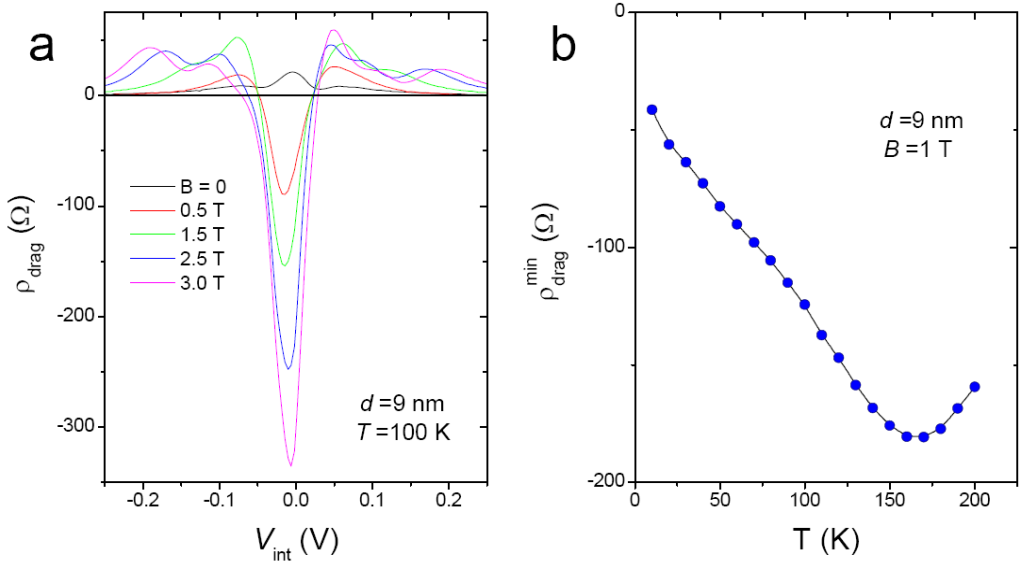

Figure S3. Strong negative drag in magnetic field. **a** – Device with $d$ =9 nm shows smaller but qualitatively the same drag as in Fig. 3b of the main text ($d$ =6nm). **b** – Magnitude of drag at the dual NP at various $T$. The negative drag is strongest at $T \approx 150$ K.

Unlike the narrow positive peak that is sample dependent, strong negative magneto-drag at the dual NP is highly reproducible from sample to sample and only its magnitude changes with *d*. To illustrate this fact, Figure S3 shows drag in a double-layer device with *d* =9 nm. Magnetic field *B* rapidly reverses the sign of $\rho_{drag}$ at zero *n* and drag's strength becomes very large, although still several times smaller than under the same conditions for the devices in Fig. 3b (*d* =6 nm). It is also worth noting that in *B* >2T drag exhibits Landau oscillations at finite *n* (Fig. S3a). Although the oscillations become more pronounced at higher *B*, we did not observe the sign change in $\rho_{drag}$ at high *n*, that is, for all LLs other than zero the drag remained positive.

To show that the $T^2$ dependence reported in Fig. 2a of the main text (device with *d* ≈6nm) is universal and persists if we further increase the interaction strength, Figure S2a plots drag's *T* dependence in the Fermi liquid regime for our thinnest spacer (*d* =1nm; *d/l* <0.1). One can see again that the drag closely follows the $T^2$ dependence, and notable deviations occur only for the upper curve at *T* >100K. The reason for the deviations is obvious: drag starts saturating and reaches a maximum value at *T* ≈$E_F$/2 ≈220K for this particular *n*. Further increase in *T* or decrease in *n* pushes the system into the Boltzmann regime where $\rho_{drag}$ decreases with increasing *T*. Note that Ref. S3 reported *T* dependence slower than quadratic, which disagrees with our experiment.

The suggested mechanism for the central peak in Fig. 3a implies that the peak should always be present, if $\delta n$ is sufficiently large or *T* sufficiently low. This is because of drag's *T* dependence, which is nonmonotonic at the dual NP and approximately varies as ∝*T* at low *T*, that is slower than the general background (see Fig. 3a). To prove that the peak is a universal feature, we have taken one of our devices that exhibited no apparent peak at the dual NP for *T* >100K and investigated it in more detail. Fig. 2b reveals that although the central peak in this device is very small, it still emerges below 70K.

## 3. Functional dependence of Coulomb drag

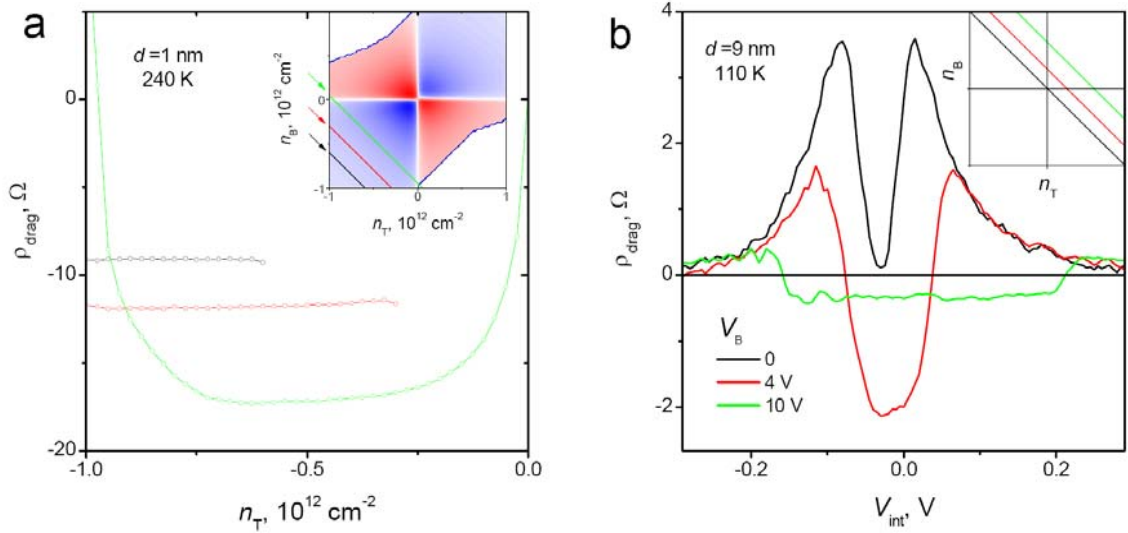

Figure S4. Functional dependence of Coulomb drag in zero *B*. Away from the NPs, the measured $\rho_{drag}$ is practically constant along the lines of constant ($n_T + n_B$). Different colors refer to different cross-sections in the 3D plot of Fig. 1c, which for convenience is shown as the inset. **b** – The same functional behavior found for *d* =9 nm. By applying $V_{int}$ we create equal concentrations $n_T$ =-$n_B$ in the two layers. Back gate voltage $V_B$ shifts the curves away from the dual NP as illustrated in the inset (no 3D plot was obtained for this device; takes more many days of continuous measurements). A long region of constant ($n_T + n_B$) is achieved for $V_B$ =10 V (green), which is sufficiently away from zero $n_T$ and zero $n_B$.

Figure 1c of the main text shows strong disagreement between the functional dependence expected in the weakly interacting regime $\rho_{drag} \propto 1/(n_T \times n_B)^{3/2}$ and our strong-coupling experiment that shows $\rho_{drag}$ to behave as $\propto 1/(n_T+n_B)^\alpha$. This is seen as flat iso-levels in Fig. 1c. To illustrate further that the experimental behavior is indeed very close to $\rho_{drag} = f(n_T+n_B)$, Fig. S4a shows several cross sections of the 3D plot from Fig. 1c along the lines with $n_T+n_B$ = constant. One can see that away from the two NPs, $\rho_{drag}$ is constant with high accuracy. Figure S4b provides another example by using a device with a relatively large $d$ =9 nm. These examples illustrate that the observed functional dependence $f(n_T+n_B)$ is a rather robust effect (also, see Fig. 4 in ref. S3).

**4. Comparison between experiment and theory in the Fermi liquid regime**

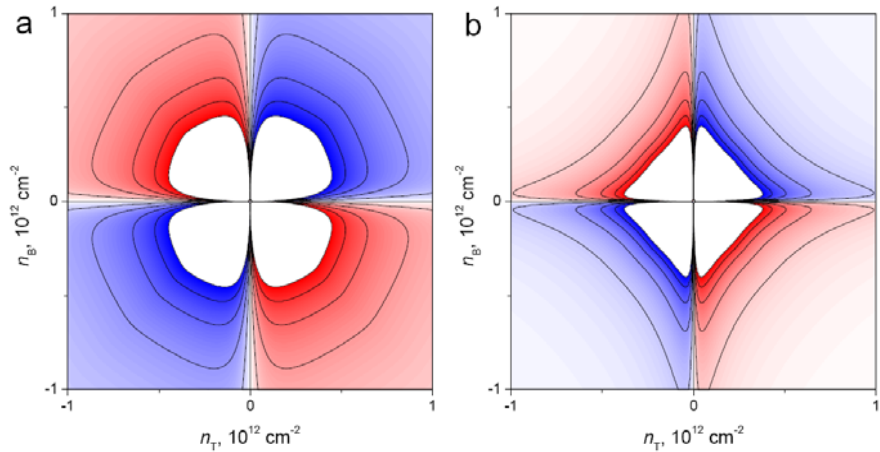

Figure S5. Predicted functional dependences of $\rho_{drag}$ for $d$ = 1 and 9 nm in **a** and **b**, respectively. These are to be compared with the experimental dependences in Figs. 1c & S4. The iso-levels in (a) are at every 5$\Omega$; the color range covers ±25$\Omega$. For (b), it is 2$\Omega$ and ±10$\Omega$, respectively. White areas: diverging theory data for $\rho_{drag}$ near the double NP are omitted.

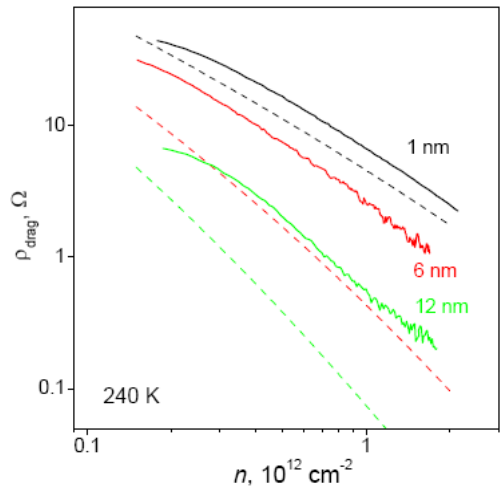

Figure S6. Solid curves are the data from Fig. 2b and the dashed curves are theory (no fitting parameters). Different $d$ are in matching colors. We do not show curves for the intermediate values of $d$ to avoid the plot's overcrowding. The bending at low $n$ for the experimental curves is due to proximity of the Boltzmann regime whereas theory [S4] assumes the Fermi liquid behavior for all $n$.

Figure S5 shows $\rho_{drag}$ calculated as a function of $n_T$ and $n_B$ in the same representation as in Fig. 1c, for $d$ =1 and 9 nm (to directly compare with devices in Figs. 1 & S4). We have used the theory presented in [S4]. One can see that for $d$ =9 nm the iso-levels are expected to be strongly curved inwards and, therefore, the functional dependence resembles that in the weakly interacting limit $\rho_{drag} = f(n_T \times n_B)$. For $d$ =1 nm (strongly-interacting limit), the functional dependence in Fig. S5 completely changes, and the iso-levels are curved in the opposite direction. The theoretical behavior is very different from that observed experimentally, which shows the robust functional dependence $f(n_T+n_B)$ for both values of $d$ (also, see Fig. 4 in ref. [S3]). The reason for this disagreement remains to be understood.

In our discussion of the observed dependence $\rho_{drag}(n)$ in Fig. 2b of the main text, we stated good agreement between the exponents $\alpha$ observed experimentally and those calculated by using the recent theory [S4]. No actual comparison is presented in the main text. This is done in Fig. S6. The theory reproduces well the experimental slopes $\alpha$ and their gradual change with $d$. The only significant difference is a factor of about 3 for the absolute value of $\rho_{drag}$. One can perhaps accommodate this missing coefficient by invoking various enhancement mechanisms [S5].

## 5. Drag in the regime of *e-h* puddles

Our observation of the large positive drag at the dual NP in zero *B* has already attracted two theoretical explanations [S6,S7] (see main text). We suggest a third one. If the central peak were negative, it would be straightforward to explain it by *e-h* puddles. Indeed, these puddles are generally believed to be due to an electrostatic potential created by charged impurities. This should then result in a spatially correlated inhomogeneity so that *e(h)* puddles in one graphene layer align with *e(h)* puddles in the other, as also noted in ref. [S7]. Therefore, an external electrostatic potential can only lead to negative Coulomb drag at the dual NP. To explain the positive drag, it requires anticorrelations in the distribution of *e-h* puddles. To this end, we speculate that charge inhomogeneity in our high-$\mu$ graphene has predominantly a non-electrostatic origin, that is, due to random strain. Random strain is known to lead to *e-h* puddles [S8], and our observation that rapid thermal cycling can significantly increase $\delta n$ without changing $\mu$ supports the model of strain-induced *e-h* puddles in our devices. Under this assumption, we are able to explain the positive peak and, qualitatively, its *T* and *n* dependences (see Fig. S7).

We have modeled the problem as follows. First, we generated a pseudo-electric potential with a finite amplitude $\delta\phi_T$ and $\delta\phi_B$ in the top and bottom layers, respectively, by using a random superposition of harmonics with wavelength *D*. Then, we used the linear response theory within the random phase approximation [S4] to calculate the induced charge densities in the two layers for a given separation *d*. For a large interlayer separation ($d \gg D$), *e-h* puddles in the two layers are uncorrelated, as expected (Fig. S7a). However, if $d \ll D$, Coulomb interactions between the layers overwhelm the strain (pseudo-electric) energy and lead to strong anticorrelations so that *e(h)* puddles in one layer appear on top of *h(e)* puddles in the other (Fig. S7b). Such a distribution of *e-h* puddles leads to positive drag. The employed approach is accurate if $\delta\phi$ is smaller than typical $E_F$ but we expect the qualitative picture for the generated puddles to be valid more generally. We have examined two cases: $\delta\phi_T = \delta\phi_B$ (graphene layers are equally strained) and $\delta\phi_B \ll \delta\phi_T$ (one layer has higher quality). In both cases, we find that puddles' distributions in the two layers become strongly anticorrelated for $d \ll D$ as illustrated in Fig. S7b.

Finally, by using the local density approximation ($D \gg l$) and the dependence $\rho_{drag}(n)$ found in ref. [S9] for the case of no puddles (zero $\delta n$), we have calculated the global drag $\langle\rho_{drag}\rangle$ by averaging the spatially varying local drag resistivity. Figure S7c shows the resulting behavior of $\langle\rho_{drag}(n)\rangle$, which is plotted in the same representation as in Fig. 3a of the main text. One can see that the model reproduces all the main features in the experimental behavior.

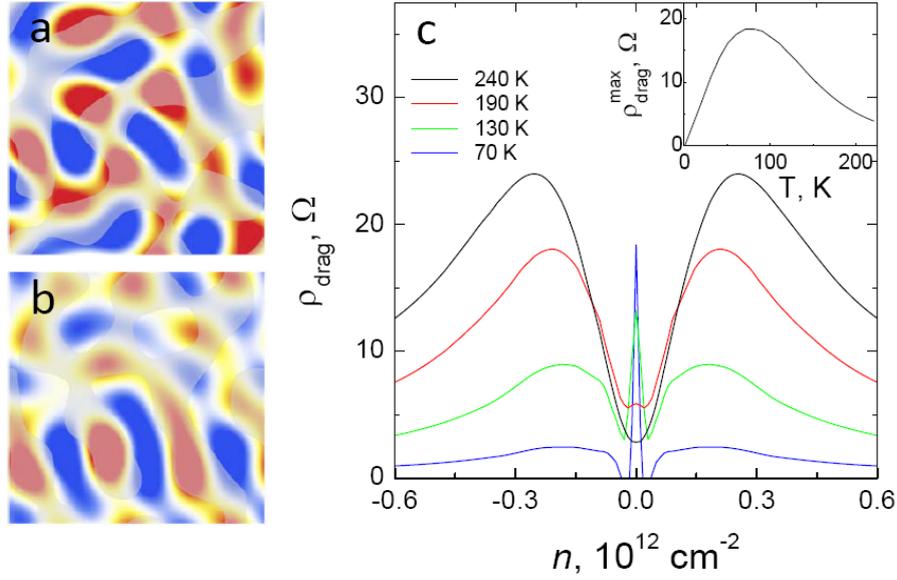

Figure S7. Modeling of Coulomb drag in the presence of *e-h* puddles. **a,b** – random puddles induced by strain in the two layers placed on top of each other; $\delta\phi_B =0.1\delta\phi_T$. The bottom layer is shown in color whereas the top one is semitransparent. Puddles are assumed to have a typical size *D* of 50 nm (see, refs. [S10,S11]). **a** – If $d \gg D$, *e-h* puddles in each layer do not talk to each other and are independent. **b** – Assuming the same initial distributions as in (a) and then bringing the layers at $d =4$ nm ($d \ll D$). One can see that electrostatic interaction overcomes strain and makes strongly (anti)correlated distributions. **c** – $\rho_{drag}(n)$ in the presence of anticorrelated *e-h* puddles ($\delta\phi_T =30$ meV; $d =9$ nm). $\delta\phi_B$ is chosen to be zero in (c) because the modeled experimental device in Fig. 3a also had a high quality bottom layer. The inset shows *T* dependence at the dual NP found in this modeling.

The nonmonotonic *T* dependence found in our modeling (inset in Fig. S7b) is related to a finite depth $\delta n$ of *e-h* puddles so that average drag becomes strongest at some temperature $T^* \approx 2E_F^*$ [S9] where $E_F^*$ corresponds to density $\delta n$ within a typical puddle. For $T > T^*$, puddles are mostly in the Boltzmann regime and the resulting drag decreases with increasing *T* (within each puddle, $\rho_{drag} \propto T^{-2}$) [S9]. At $T < T^*$, the puddles enter the Fermi liquid regime, and $\langle \rho_{drag} \rangle$ again decreases with decreasing *T* (locally, $\rho_{drag} \propto T^2$). This explains why strongest peaks occur in devices with largest $\delta n$.

Our model also reproduces the linear *T* dependence found experimentally at low *T* (cf. insets in Fig. 3a and S7c). This dependence effectively arises from averaging over puddles of different depth. Indeed, let us assume that the Fermi energies in both layers are determined by the same random and spatially varying quantity $\delta\phi(r)$. This is the case of $\delta\phi_T \gg \delta\phi_T$ as both distributions are determined by charge inhomogeneity in the top layer. According to ref. [S9], $\rho_{drag}$ is a function of $\delta\phi/T$ with the asymptotes $x^2$ and $1/x^2$ for $x \ll 1$ and $x \gg 1$, respectively. Then, $\langle \rho_{drag} \rangle(T) = \int d(\delta\phi) g(\delta\phi) \rho_{drag}(\delta\phi/T)$ where $g(\delta\phi)$ is the probability density for $\delta\phi$. Since $\delta\phi$ is modeled as a sum of a large number of harmonics with randomly distributed phases, it should be Gaussian and, thus, *g* approaches a finite value for $\delta\phi \to 0$. Accordingly, for small *T*, we can write $\langle \rho_{drag} \rangle(T) = g(0) T \int d(x) \rho_{drag}(x) \propto T$ which explains the *T* dependence shown in Fig. S7c. Note that the alternative explanation [S7] for the observed peak at the dual NP predicts much stronger dependence ($\propto T^4$) in the low *T* limit. Unfortunately, we cannot definitively exclude a stronger low-*T* behavior because mesoscopic fluctuations limit our accuracy at low *T* (Fig. 3a).

The central peak in $\langle \rho_{drag} \rangle$ appears in our modeling due to stronger interlayer correlations as the system approaches the dual NP. This can be understood as follows. Average drag depends on the correlation function $C(n) = \langle n_T(r) n_B(r) \rangle / \langle n_T(r) \rangle \langle n_B(r) \rangle$ that is given by spatially varying densities $n_{T,B}(r)$ in the two layers. For $n \gg \delta n$, the inhomogeneity induced by strain is screened, both layer are filled uniformly ($n_T = -n_B$) and $C \to 1$. At zero $n$ and for strongly-interacting layers ($d \ll D$), correlations in the distributions of $e$-$h$ puddles are also strong (Fig. S7b) and, again, we get $C \approx 1$. However, $C(n)$ is not a constant as a function of $n$ and, as found in our modeling, exhibits a dip at $n < \delta n$. This is due to increased screening with increasing $n$ which reduces the interlayer correlations. This nonmonotonic behavior of $C(n)$ effectively means that $C(n)$ exhibits a relatively sharp rise near zero $n$. $<\rho_{drag}>$ is given by a convolution of $C(n)$ with the functional dependence $\rho_{drag}(n)$ found in ref. [S9], and the rise in $C(n)$ leads to the central peak at the dual NP. Our modeling also shows that the dip in $C$ and, hence, the peak in $<\rho_{drag}>$ are strongest if one layer has shallow puddles ($\delta\phi_B \ll \delta\phi_T$), which agrees with the fact that the strongest peak was observed in a device with a high-quality bottom layer.

Although our model of strain-induced $e$-$h$ puddles qualitatively explains the behavior of drag at the dual NP in zero $B$, we note that $D$ is expected to be ~50 nm for graphene on hBN [S10,S11] and, therefore, puddles contain only a few charge carriers, that is, $D \sim l$. Analysis for Coulomb drag for such a microscopically inhomogeneous case is a complex problem beyond the scope of the present paper.

**Supplementary references**